\documentstyle[12pt]{article}

\newcommand{\be}{\begin{equation}}
\newcommand{\ee}{\end{equation}}
\newcommand{\bea}{\begin{eqnarray}}
\newcommand{\eea}{\end{eqnarray}}

 at 14.4truept
 at 14.4truept

\def\pa{\partial}
\def\ker{{{\rm Ker}({\rm ad\,} \Lambda)}}
\def\kerr{{{\rm Ker}({\rm ad\,} \Lambda_0)}}
\def\kerrr{{{\rm Ker}({\rm ad\,} I_-)}}
\def\kerh{{{\rm Ker}({\rm ad\,} E)}}
\def\im{{{\rm Im}({\rm ad\,} \Lambda) }}
\def\H{{\cal H}}

\def\C{{\cal C}}
\def\A{{\cal A}}

\def\M{{\cal M}}

\def\S{{\cal S}}
\def\W{{\cal W}}
\def\G{{\cal G}}
\def\L{{\cal L}}
\def\M{{\cal M}}

\def\N{{\cal N}}

\def\1{{ 1}}

\begin{document}

\renewcommand{\thefootnote}{\fnsymbol{footnote}}
\fnsymbol{footnote}

\rightline{ SWAT-95-89\break}
\rightline{INS-Rep.-1114\break}
\rightline{ September 1995\break}

\begin{center} {\Large \bf KdV type systems  and $\cal W$-algebras\\

in the Drinfeld-Sokolov
approach}\protect\footnote{Talk given
at the Marseille Conference  on $\cal W$-Symmetry, July 1995.}

\end{center}

\vspace{.15in}

\begin{center}
L\'aszl\'o~Feh\'er\\

\vspace{0.1in}

{\sl Department of Physics, University of Swansea,
Singleton Park, Swansea SA2 8PP, UK}\\

and\\

{\sl Institute for Nuclear Study, University of Tokyo, Tanashi,
Tokyo 188, Japan}\protect\footnote{Present address.  On leave from
Theoretical Physics  Department  of Szeged University, Szeged,
Hungary.}\\

\end{center}

\vspace{.15in}

{\parindent=25pt
\narrower\smallskip\noindent

\small
\noindent
{\bf Abstract.}\quad
The generalized Drinfeld-Sokolov construction of KdV  systems
is reviewed in the case of an arbitrary affine Lie algebra
paying particular attention to Hamiltonian aspects and $\W$-algebras.
Some  extensions of known results as well as a new
interpretation of the construction
are also presented.
}

\normalsize

\vspace{0.15in}
\section{Introduction}
\renewcommand{\thefootnote}{\arabic{footnote}}
\setcounter{footnote}{0}

In order to set  the context of this talk,
let us recall  (see e.g.~the reviews in \cite{Di,BS})
that the so called
second Hamiltonian structure
of the KdV hierarchy, for which the  Lax operator is
$L=\pa_x^2 + u(x)$, is the Poisson bracket version of the
Virasoro algebra.
The simplest generalized KdV hierarchies  are
the $n$-KdV hierarchies, which are  based on  Lax operators  of the form
\begin{equation}
L=\pa_x^n+u_2(x)\pa^{n-2}+\cdots +u_n(x),
\label{1.1}\end{equation}
and whose  second (Gelfand-Dickey) Hamiltonian structure
defines the $W_n$-algebra.

Drinfeld and Sokolov \cite{DS} constructed a set of generalizations
of the $n$-KdV hierarchy using a Lie algebraic framework.
In their construction the phase space of a generalized
KdV system  appears as the  space of orbits  --- reduced phase space ---
of a certain gauge group in a manifold.
The manifold in question  consists of first order differential
operators
\begin{equation}
\L=\pa_x + j(x) + \Lambda,
\label{1.2}\end{equation}
where $\Lambda$ is a regular semisimple element of an affine
Lie algebra $\A$ with principal grade one.
The form of $j(x)$ and that of the gauge group are  specified
with the aid of the principal grading of $\A$ together with a standard
grading of $\A$.
The system is Hamiltonian with respect to a  Poisson bracket
that may be identified as the $\W$-algebra corresponding to
the principal $sl_2$ embedding in the finite dimensional
reductive Lie algebra $\A_0\subset \A$, where $\A_0$  has zero grade
in the relevant standard grading of $\A$.
The  simplest examples   are obtained by taking
$\A=\G\otimes {\bf C}[\lambda,\lambda^{-1}]$ for $\G$ a
finite dimensional simple Lie algebra  and
choosing the standard
grading to be the homogeneous grading.
Then $\A_0=\G$ and the second Hamiltonian structure is
the  well-known principal (Casimir) $\W$-algebra associated
to $\G$.  The $W_n$-algebra is recovered for $\G=sl_n$.
In the case of  $\A$ a twisted loop algebra
the Drinfeld-Sokolov KdV systems  have only a single
Hamiltonian structure in general, which is still often
referred to as the ``second'' one.

Drinfeld and Sokolov also constructed modified KdV systems
related to their KdV systems by generalized Miura maps. There is such
a modified KdV system for every affine Lie algebra, having
a phase space of elements of the form
\begin{equation}
\L_{\rm mod-KdV}=\pa_x + \theta(x) + \Lambda,
\qquad \theta(x)\in \A^0,
\label{1.3}\end{equation}
where $\A^0$ is the abelian subalgebra of $\A$ given by the
elements of principal grade zero.

Generalized {\em Drinfeld-Sokolov reduction}
has received a lot of interest in the last few years.
On the one hand,
this term has been used  to refer to Hamiltonian reductions of
current algebras based on simple Lie algebras
to  $\W$-algebras that possess a  finite
generating set consisting  of a Virasoro field and conformal
tensors.
The main success here has been the construction of a
classical $\W$-algebra to every $sl_2$ embedding into a simple Lie
algebra $\G$ \cite{bais,rep}
and the construction of the corresponding quantum
$\W$-algebra by the BRST technique \cite{dBT,ST}.
We also have some classification results \cite{bowwatts,fehort,FORT}
pointing to a  distinguished role of these $\W$-algebras among 
classical
extended conformal algebras based on finitely generated,
freely generated differential rings.

In the literature  the  $\W_\S^\G$-algebra \cite{bais,rep}
is usually considered
for an $sl_2$ embedding $\S\subset \G$ into a {\em simple} Lie
algebra $\G$, but the construction extends in an obvious way
to any {\em reductive} Lie algebra.
A reductive Lie algebra, like $gl_n$,
decomposes into the direct sum of a  semisimple and a central part,
and any $sl_2$ embedding is contained in the semisimple factor.

The other sense in which the notion of generalized Drinfeld-Sokolov
reduction has been used in the literature concerns the construction
of KdV type systems (see refs.~[11--20]).

The common feature of Drinfeld-Sokolov reductions in the
$\W$-algebra context and in the KdV context is that the existence
of ``DS gauges''  is required in order  to obtain a finite
set of independent differential polynomial invariants with respect to
a non-trivial gauge group.
In the KdV context these invariants are the generalized KdV fields
themselves.
The ``second''  Poisson bracket algebra is not necessarily a
$\W$-algebra in all generalized KdV systems.
Conversely, according to  present knowledge,
not all classical $\W$-algebras resulting from
generalized Drinfeld-Sokolov reductions support KdV type hierarchies.

In the rest of this talk we review the construction
of modified KdV  and KdV type systems,
and also review what is known about the classification of such systems
and their relationship to $\W$-algebras.
To proceed in order of increasing complexity and for logical reasons,
it will be convenient to first recall the construction of
systems of modified KdV type.

\medskip
\section{Generalized modified KdV systems}
\setcounter{equation}{0}

The subsequent construction is  due to Wilson \cite{Wi} with
elaborations and extensions contributed  by McIntosh \cite{McI} and
de Groot et al \cite{Prin1}.
It associates  a hierarchy of ``modified KdV type'' to any triplet
\begin{equation}
\left(\A, d_\sigma, \Lambda\right),
\label{2.1}\end{equation}
where $\A$ is an affine Lie algebra with vanishing centre,
$d_\sigma$ is a $\bf Z$-grading of $\A$, and $\Lambda$ is a semisimple
element of $\A$ which is homogeneous of $d_\sigma$-grade $k>0$.

The affine Lie algebra $\A$ can be realized as a twisted loop algebra
\begin{equation}
\A=\ell(\G,\mu)\subset \G\otimes {\bf C}[\lambda,\lambda^{-1}]
\label{2.2}\end{equation}
for a finite dimensional complex simple Lie algebra $\G$ and
an automorphism $\mu$ of $\G$ of finite order.
To investigate concrete systems it is often
convenient to choose such a realization of $\A$,
which means fixing $\mu$, but for our purpose  it is better to define
$\A$
abstractly \cite{kac} in terms of  its Chevalley generators belonging to
the simple roots,
\begin{equation}
e_i,\quad f_i,\quad h_i,\quad i=0,1,\ldots,r.
\label{2.3}\end{equation}
These  satisfy the relations
\begin{equation}
[e_i, f_j]=\delta_{i,j} h_i,
\quad
[h_i, e_j]=K_{j,i} e_j,
\quad
[h_i, f_j]=-K_{j,i} f_j,
\label{2.4}\end{equation}
where $K_{j,i}$ is the Cartan matrix of $\A$, together with
the Serre relations
and a relation expressing the vanishing of the centre,
\begin{equation}
\sum_{i=0}^r n_i h_i =0
\label{2.5}\end{equation}
with some non-negative integers $n_i$.
Without loss of generality,
the derivation $d_\sigma:\A\rightarrow \A$, whose eigenspaces
$\A^n$ fix a grading $\A=\oplus_n \A^n$ of $\A$, can be defined by
\begin{equation}
d_\sigma(e_i)=\sigma_i e_i,
\quad
d_\sigma(f_i)=-\sigma_i f_i,
\quad
d_\sigma(h_i)=0,
\label{2.6}\end{equation}
where $(\sigma_0, \sigma_1,\ldots,\sigma_r)$ is a set
of relatively prime non-negative
integers\footnote{The grading $d_\sigma$ is called a
{\em standard} grading if $\sigma_i=\delta_{i,j}$ for some
$j\in\{0,1,\ldots,r\}$.}.
The assumption that
$\Lambda\in \A$ is a graded semisimple element
means that it defines a direct sum decomposition
\begin{equation}
\A=\ker +\im,
\qquad
\ker\cap\im =\{ 0\},
\label{2.7}\end{equation}
and it satisfies $d_\sigma(\Lambda)=k\Lambda$ for some
integer $k>0$.

By definition, the phase space of the mod-KdV hierarchy associated
to the triplet $(\A, d_\sigma, \Lambda)$ is the manifold $\Theta$
of first order differential operators given by
\begin{equation}
\Theta:=\left\{ \L=\pa_x + \theta(x) +\Lambda\,\vert\,
\theta(x)\in \A^{<k} \cap \A^{\geq 0}\,\right\}.
\label{2.8}\end{equation}
We use the notation $\A^{<k}=\oplus_{n<k} \A^n$ etc.
Since $\A^{<k}\cap \A^{\geq 0}$ is a finite dimensional space,
the mod-KdV field $\theta(x)$ encompasses a finite
number of compex valued fields depending on the one-dimensional
space variable $x$.
There is a natural family of compatible evolution equations on $\Theta$,
whose members are labelled by  the  graded basis elements of the space
\begin{equation}
\C(\Lambda):=\left({\rm Cent}\, \ker\right)^{\geq 0},
\label{2.9}\end{equation}
which is the positively graded part of the centre of the Lie
algebra $\ker$.
In order to construct these equations,
one makes use of  the transformation
\begin{equation}
\L=(\pa_x + \theta(x) +\Lambda)\mapsto e^{{\rm ad} F}(\L)=
\pa_x + h(x) + \Lambda,
\label{2.10}\end{equation}
for $F(x)$ and $h(x)$ being infinite series
\begin{equation}
F(x)\in \left(\im\right)^{<0},
\qquad
h(x)\in \left(\ker\right)^{<k}.
\label{2.11}\end{equation}

\medskip
\noindent{\bf Basic lemma I.}
{\em For arbitrarily given  $\theta(x)$,
the above equations have a unique solution  $F(x)=F(\theta(x))$,
$h(x)=h(\theta(x))$. The components of $F(\theta(x))$
and $h(\theta(x))$ are differential polynomials
in the components of $\theta(x)$.}
\medskip

Thanks to the  lemma, which goes back to Drinfeld and Sokolov
(see also \cite{McI,Prin1}), for any element $b\in \C(\Lambda)$
one can define
 \begin{equation}
A(b,\theta):=e^{-{\rm ad} F(\theta) } (b),
\label{2.12}\end{equation}
and   $A(b,\theta)$
is a differential polynomial in $\theta$.
Using the grading, $A(b,\theta)$ is decomposed as
\begin{equation}
A(b,\theta)=  A(b,\theta)^{\geq 0}+A(b,\theta)^{<0} .
\label{2.13}\end{equation}
The flow equation associated to $b\in \C(\Lambda)$ is given by
the  vector field ${\pa \over \pa T_b}$ on $\Theta$:
\begin{equation}
{\pa \theta  \over \pa T_b }:=\left[ A(b,\theta )^{\geq 0}, \L\right]
=-\left[A(b,\theta)^{<0}, \L\right].
\label{2.14}\end{equation}
Equivalently,
\begin{equation}
\left[ {\pa \over \pa T_b} - A(b,\theta(x))^{\geq 0}\,,\,
\pa_x +\theta(x) +\Lambda \right]=0.
\label{2.15}\end{equation}
The flows corresponding to different elements
of $\C(\Lambda)$ commute,
\begin{equation}
\left[ {\pa \over \pa T_b}\,,\, {\pa \over \pa T_{b'}}\right]=0
\qquad \forall\,  b,b'\in \C(\Lambda).
\label{2.16}\end{equation}

To understand the Hamiltonian interpretation of the above flows,
it is convenient to first introduce the large space
\begin{equation}
\tilde \A:=\{ \L=\pa_x + J(x)\,\vert\, J(x)\in \A\,\}.
\label{2.17}\end{equation}
Assuming  appropriate smoothness and
boundary conditions (e.g.~periodic boundary condition)
on the field $J(x)$,
the following formula defines a Poisson bracket on $\tilde \A$,
\begin{equation}
\{ f, g\}_\sigma (J):=\int dx\,
\left\langle J,
\left[ \left( {\delta f\over \delta J}\right)^{\geq 0} ,
 \left( {\delta g\over \delta J}\right)^{\geq 0}\right]
-\left[ \left( {\delta f\over \delta J}\right)^{< 0},
 \left( {\delta g\over \delta J}\right)^{< 0}\right]
\right\rangle
-\left\langle \left( {\delta f\over \delta J}\right)^{0},
\pa_x \left( {\delta g\over \delta J}\right)^{0}\right\rangle,
\label{2.18}\end{equation}
where $\langle\ ,\ \rangle$ denotes the non-degenerate,
invariant, symmetric bilinear form on $\A$, which is unique up to
a constant, and
the functional derivative ${\delta f\over \delta J(x)}\in \A$ is defined
in the natural way.

In fact \cite{RSTS},  $\Theta\subset \tilde \A$ is a
{\em Poisson submanifold\,}\footnote{
Remember that one can trivially restrict the Poisson bracket to
a Poisson submanifold which
is invariant with respect to any hamiltonian flow by definition.}
of $(\tilde\A, \{\ ,\ \}_\sigma)$.
The flow in (\ref{2.14}) is hamiltonian
with respect to the induced Poisson bracket on $\Theta$ and the
Hamilton function  $H_b(\theta)$ given by
\begin{equation}
H_b(\theta)=\int dx\, \langle b, h(\theta(x))\rangle.
\label{2.19}\end{equation}

Drinfeld and Sokolov \cite{DS} considered the systems for which
$d_\sigma$ is the principal grading, given by
 $\sigma_0=\sigma_1=\cdots =\sigma_r=1$,
and $\Lambda$ is the (essentially unique) grade one regular
semisimple element, whose centralizer --- disregarding
the central extension --- is the principal Heisenberg
subalgebra of $\A$.
Wilson \cite{Wi}  proposed the above construction for a general grading
and $\Lambda$ a {\em grade one} semisimple element, which
is clearly the nicest case in the sense that the
number of fields is the smallest for a given grading $d_\sigma$.
This case, including a partial classification, was elaborated
by McIntosh \cite{McI}.
The extension for $\Lambda$ having arbitrary positive grade
is due to de Groot et al \cite{Prin1,Prin2,Prin3}.

\medskip
\section{Construction of KdV type systems}
\setcounter{equation}{0}

The construction of KdV type systems is much more subtle than
that of the mod-KdV type systems  since
it requires the existence of a finitely generated, freely generated
ring of differential polynomial invariants
with respect to a non-trivial gauge group.
As a consequence, it is not necessarily the case
that  every system of mod-KdV type has a Miura map which relates
it to a system of KdV type\footnote{Our notion
of {\em KdV type system} is defined by the subsequent
construction.
In the literature these systems are often
called ``partially modified KdV  systems'' reserving the term ``KdV
system''
for the special case of $d_\tau$ in (\ref{3.1}) being a standard grading
of $\A$.}.
The existence of the data given below
is  a  sufficient condition for the existence
of a Miura map.

The set of data to which a  ``KdV type system'' can be
associated \cite{McI,Prin1}
is a quadruplet
\begin{equation}
(\A, d_\sigma, \Lambda, d_\tau)
\label{3.1}\end{equation}
subject to certain conditions that include those required previously for
the triplet $(\A,d_\sigma,\Lambda)$.
That is  $\Lambda\in \A$ is a semisimple element of $d_\sigma$-grade 
$k>0$.
Here $d_\tau$ is another ${\bf Z}$-grading of
$\A$, which is {\em compatible} with $d_\sigma$ in the sense that
there exist Chevalley generators of $\A$ in terms of which $d_\sigma$
is given as in (\ref{2.6}) and $d_\tau$ is given similarly in the
same basis,
\begin{equation}
d_\tau(e_i)=\tau_i e_i,
\quad
d_\tau(f_i)=-\tau_i f_i,
\quad
d_\tau(h_i)=0,
\label{3.2}\end{equation}
where $(\tau_0, \tau_1,\ldots,\tau_r)$ is a set
of relatively prime non-negative integers.
It is also assumed  that
\begin{equation}
\sigma_i=0\Longrightarrow \tau_i=0.
\label{3.3}\end{equation}
The two compatible gradings together define a bi-grading of $\A$,
\begin{equation}
\A=\oplus_{m,n} \A_m^n,
\qquad
\A_m^n:=\{\,X\in \A\,\vert\, d_\sigma(X)=nX,
\,\,\, d_\tau(X)=mX\,\},
\label{3.4}\end{equation}
so that $\A_m^n$ has $d_\sigma$-grade $n$ and $d_\tau$-grade $m$.
The relation in (\ref{3.3}) together with
$\tau_i\sigma_i\geq 0$ means that $d_\tau$ is ``coarser'' than
$d_\sigma$,
\begin{equation}
\label{3.5}\A^0\subseteq \A_0.
\end{equation}
In \cite{Prin1} the notation $\tau\preceq \sigma$ was used to denote
the relation between the two gradings, and the following
useful consequences of this relation were also noted,
\begin{equation}
\A_{>0}\subseteq \A^{>0},
\quad
\A_{<0}\subseteq \A^{<0},
\qquad
\A^{>0}\subset \A_{\geq 0},
\quad
\A^{<0}\subset \A_{\leq 0}.
\label{3.5+}\end{equation}
Here and below superscripts denote $d_\sigma$ grades and
subcripts $d_\tau$ grades as in (\ref{3.4}).
Finally, we require the  ``non-degeneracy condition'' on the
quadruplet given by
\begin{equation}
\ker\cap \A_0^{<0}=\{0\}.
\label{3.6}\end{equation}
This is a non-trivial condition if $\A_0^{<0}\neq \{0\}$,
i.e.~if $\A_0\neq \A^0$, which we
also assume.

By definition, the phase space of the KdV system associated
to the quadruplet $(\A,d_\sigma, \Lambda, d_\tau)$ is the factor space
\begin{equation}
\M_{\rm red}=\M_c/ {\cal N},
\label{3.7}\end{equation}
where
\begin{equation}
\M_c=\left\{ \L=\pa_x + j(x) +\Lambda\,\vert\,
j(x)\in \A_{\geq 0}^{<k}=\A^{<k}\cap\A_{\geq 0}\,\right\}
\label{3.8}\end{equation}
and $\N$ is the group of ``gauge transformations'' $e^\alpha$ acting
on $\M_c$ according to
\begin{equation}
e^\alpha: \L\mapsto e^{{\rm ad} \alpha}\left(\L\right)=
e^\alpha \L e^{-\alpha},
\qquad
\L\in \M_c,
\quad
\alpha(x)\in \A_0^{<0}.
\label{3.9}\end{equation}
That is the Lie algebra of  $\N$ is given by
$\tilde \A_0^{<0}:=\{ \alpha\,\vert\, \alpha(x)\in \A_0^{<0}\}$
(with some smoothness and boundary conditions for $\alpha(x)$).

Let $V\subset \A_{\geq 0}^{<k}$ be a $d_\sigma$-graded vector
space defining a direct sum decomposition
\begin{equation}
\A_{\geq 0}^{<k}=\left[\Lambda, \A_0^{<0}\right] + V.
\label{3.10}\end{equation}
Define   $\M_V\subset \M_c$  by
\begin{equation}
\M_V:=\left\{ \L=\pa_x + j_V(x) +\Lambda\,\vert\,
j_V(x)\in V\,\right\}.
\label{3.11}\end{equation}
Due to the non-degeneracy condition  (\ref{3.6}) and the grading
assumptions, the action of $\N$ on $\M_c$ is a free action and
we have

\medskip
\noindent{\bf Basic lemma II.}
{\em
The submanifold $\M_V\subset \M_c$ is a global cross section
of the gauge orbits defining  a one-to-one model of
$\M_{\rm red}=\M_c/\N$.
When regarded as functions on $\M_c$,
the components of $j_V(x)=j_V(j(x))$ are differential polynomials, which
thus provide a free generating set of the ring $\cal R$
of gauge invariant differential polynomials on $\M_c$.}
\medskip

Since this  lemma also  goes back to
Drinfeld and Sokolov,
we refer to $\M_V$ as a {\em DS gauge} (see also \cite{rep,FORT}).
The construction of commuting local flows on $\M_c$ given
in \cite{McI,Prin1}
is  based on the variant of ``Basic lemma I'' which says that for any
$\L=(\pa_x+j+\Lambda)\in \M_c$
there is a unique differential polynomial $F(j(x))\in
\left(\im\right)^{<0}$
such that
\begin{equation}
e^{{\rm ad} F(j)}(\L)=\pa_x + h(j) +\Lambda,
\quad\hbox{with}\quad
h(j(x))\in \left(\ker\right)^{<k}.
\label{3.12}\end{equation}
Using this, for any $b\in \C(\Lambda)$ in (\ref{2.9})
one can define the commuting vector fields
$\pa \over \pa t_b$ on $\M_c$ by
\begin{equation}
{\pa j  \over \pa t_b }=\left[ A(b,j )_{\geq 0}, \L\right]
=-\left[A(b,j)_{<0}, \L\right],
\qquad
\L=\pa_x + j + \Lambda,
\label{3.13}\end{equation}
where $A(b,j)=e^{-{\rm ad} F(j) } (b)$ and the splitting
$A(b,j)=A(b,j)_{<0} + A(b,j)_{\geq 0}$
is now given by  $d_\tau$.

It can be  shown that the equations in (\ref{3.13})
have a gauge invariant meaning.
Algebraically, this means that any $\pa \over \pa t_b$ gives rise
to an evolutional derivation of the ring $\cal R$ of
$\N$-invariant  differential
polynomials  in $j$.
Geometrically, letting
$\pi: \M_c\rightarrow \M_c/\N\simeq \M_V$ denote the natural
mapping, the statement is that the vector field $\pa\over \pa t_b$
on $\M_c$ has a well-defined projection
$\pi_*\left({\pa \over \pa t_b}\right)$ on
 $\M_V$.
The flow induced on $\M_V$ has an equation  of the form
\begin{equation}
\pi_*\left({\pa \over \pa t_b}\right) j_V =
\left[ A(b,j_V )_{\geq 0}+\eta(j_V) , \pa_x + j_V +\Lambda \right],
\label{3.14}\end{equation}
where $\eta(j_V(x))\in \A_0^{<0}$ is a uniquely determined  differential
polynomial in $j_V$.

It can be also shown that the vector field  in  (\ref{3.14})
is hamiltonian with respect to the Hamilton function
\begin{equation}
H_b(j_V)=\int dx\, \langle b, h(j_V(x))\rangle,
\label{3.15}\end{equation}
where $h(j)$ is defined in (\ref{3.12}),
and the Poisson bracket $\{\ ,\ \}_2$ on $\M_V$ is given as follows.
In order to define $\{\ ,\ \}_2$ we identify the
space of functions on $\M_V\simeq \M_c/\N$ with the space of
 gauge invariant functions on $\M_c$.
Then, for arbitrary such funcions $f$, $g$, we have
\begin{equation}
\{ f, g\}_2 (j):=\int dx\,
\left\langle j_0+\Lambda_0,
\left[ {\delta f\over \delta j_0} ,
 {\delta g\over \delta j_0} \right]\right\rangle
-\left\langle  {\delta f\over \delta j_0},
\pa_x {\delta g\over \delta j_0}\right\rangle
-\left\langle j_{>0}+\Lambda_{>0},
\left[ {\delta f\over \delta j_{>0}},
 {\delta g\over \delta j_{>0}} \right]\right\rangle,
\label{3.16}\end{equation}
where ${\delta f\over \delta j_0(x)}\in \A_0$,
${\delta f\over \delta j_{>0}(x)}\in \A_{<0}$ are defined
with the aid of  the scalar product $\langle\ ,\ \rangle$ on $\A$ and
the decomposition $j=j_0 + j_{>0}$ of
$j(x)\in \A_{\geq 0}^{<k}$
for which $j_0(x)\in \A_0$, $j_{>0}(x)\in \A_{>0}$,
 $\Lambda=\Lambda_0+\Lambda_{>0}$.

If $d_\tau=d_\sigma$ then
the KdV system constructed above
using the  quadruplet $(\A, d_\sigma, \Lambda, d_\tau)$
becomes the mod-KdV
system of the previous section based on the
triplet $(\A, \Lambda, d_\sigma)$.
In general $\Theta\subset \M_c$.
If the gauge group $\N$ is non-trivial
the canonical mapping
\begin{equation}
\pi\vert_{\Theta} : \Theta \rightarrow \M_V
\label{3.17}\end{equation}
provides a generalized Miura map --- that is {\it a local Poissson
map having  a  non-local, non single-valued inverse} --- from
the modified KdV system $(\A, d_\sigma, \Lambda)$ to the
 KdV system $(\A, d_\sigma, \Lambda, d_\tau)$.

The above construction of KdV type systems was proposed by
de Groot et al in \cite{Prin1,Prin2,Prin3}.
Independently,  the same  construction was proposed
earlier by McIntosh in \cite{McI} using more restrictive conditions on
the quadruplet $(\A, d_\sigma, \Lambda, d_\tau)$.
In  \cite{McI} it was assumed that $\Lambda$ has $d_\sigma$-grade one
and  $d_\tau$-decomposition of the form $\Lambda=\Lambda_0+\Lambda_1$.
The ``second'' Hamiltonian structure (\ref{3.16})
was described  in \cite{Prin2} for the case
$\A=\G\otimes {\bf C}[\lambda, \lambda^{-1}]$.

In the above exposition we took  the logical path
whereby one first constructs the evolution equations
by an algebraic method
and then, almost as an afterthought, one finds  their hamiltonian
properties by  direct computations.
This was the approach taken in  refs.~\cite{DS,Wi,McI,Prin1}.
There is another approach,
advocated in special cases (with grade one $\Lambda$)
 in \cite{FHM} where the hamiltonian
properties are clear from the very
beginning and there is essentially nothing to prove with regard
to the commutativity of the flows or the Jacobi identity of the Poisson
bracket.
In the next section  this  phase space reduction
approach will be explained concentrating on the origin of
formula ({\ref{3.16}) for the ``second'' Poisson bracket.

For a fixed KdV system $(\A,d_\sigma, \Lambda, d_\tau)$,
the subsequent  interpretation relies on the Poisson
bracket $\{\  ,\  \}_\tau$ on the space $\tilde \A$ in ({\ref{2.17}),
which is given analogously to ({\ref{2.18}) by the formula
\begin{equation}
\{ f, g\}_\tau (J):=\int dx\,
\left\langle J,
\left[ \left( {\delta f\over \delta J}\right)_{\geq 0} ,
 \left( {\delta g\over \delta J}\right)_{\geq 0}\right]
-\left[ \left( {\delta f\over \delta J}\right)_{< 0},
 \left( {\delta g\over \delta J}\right)_{< 0}\right]
\right\rangle
-\left\langle \left( {\delta f\over \delta J}\right)_{0},
\pa_x \left( {\delta g\over \delta J}\right)_{0}\right\rangle,
\label{3.18}\end{equation}
where the splitting of $\delta f\over \delta J$ is defined by $d_\tau$.
It will be useful to decompose  $\Lambda$ as
\begin{equation}
\Lambda =\Lambda_0 + \Lambda_{>0},
\qquad
\Lambda_0\in \A_0,\,\,\,\Lambda_{>0}\in \A_{>0},
\label{3.19}\end{equation}
by means of the $d_\tau$ grading, since the two components
play different roles.
We introduce the component  $\Lambda_{>0}$ by defining
the submanifold $\M\subset \tilde \A$,
\begin{equation}
\M:=\left\{ \L=\pa_x + J_0(x) +J_{>0}^{<k}(x)+\Lambda_{>0}\,\vert\,
J_0(x)\in\A_0,\,\,\, J_{>0}^{<k}(x)\in  \A^{<k} \cap \A_{>0}\,\right\}.
\label{3.20}\end{equation}
It is important to notice
that $\M$ is a
{\em Poisson submanifold\,}\footnote{This was pointed out to me by
J.L.~Miramontes. See also \cite{MiPo}.}
of $(\tilde \A, \{\  ,\  \}_\tau)$.
To see this we first write down  the Hamiltonian vector field
$X(J):=\{ J, f\}_\tau$ generated by an arbitrary function $f$ on
$\tilde \A$.  Using the $d_\tau$-grading and the notation
$Y(J):={\delta f\over \delta J}$, from (\ref{3.18}) we obtain
\begin{eqnarray}
&&X(J)_{<0}=\left[ Y(J)_{\geq 0}, J_{<0}\right]_{<0},\nonumber\\
&& X(J)_{0\phantom{<}}=\left[ Y(J)_0, J_0\right]-\pa_x Y(J)_0
+\left[ Y(J)_{>0}, J_{<0}\right]_0,\nonumber\\
&& X(J)_{>0}=-[Y(J)_{<0}, J_{>0}]_{>0}.
\label{3.21}\end{eqnarray}
Inserting $J=(J_0 + J_{>0}^{<k} + \Lambda_{>0})$ into (\ref{3.21}),
it follows ---
since $Y(J)_{<0} \in \A_{<0}\subseteq\A^{<0}$ by (\ref{3.5+}) ---
that the restriction of $X(J)$ to $\M\subset \tilde \A$ is
tangent to $\M$, i.e.~$X(J=J_0+J_{>0}^{<k}+\Lambda_0)\in
\A_0+\A_{>0}^{<k}$.
This proves that $\M$ is a Poisson submanifold.
The hamiltonian interpretation of
constraining $\M$ to $\M_c$ and that of the gauge group will be given
shortly.

\medskip
\section{KdV construction as  phase space reduction}
\setcounter{equation}{0}

In the standard case of Drinfeld and Sokolov there is a well-known
interpretation \cite{DS} of the  ``second'' Poisson
bracket of the generalized KdV systems as the reduced Poisson bracket
obtained by imposing first class constraints on the manifold $\M$
and factoring by the gauge group generated by those constraints.
However, this interpretation is not
tenable in the general case, since the action of $\N$ is not generated
by the constraints defining $\M_c\subset \M$ as is easily seen.
In order to see what happens,  a general phase space
reduction procedure will be described next, which includes
the KdV construction as a particular example.

Let $(M, \{\ ,\ \})$ be a Poisson manifold.
(For ease of understanding, one might wish to think of $M$ as being
finite
dimensional.)
Suppose that we have a non-trivial left action of a
connected Lie group  $N_0$ on $M$
which is a Hamiltonian action with respect to $\{\ ,\ \}$.
This means that the infinitesimal action of the Lie algebra
of $N_0$ on $M$ is given by the Hamiltonian vector fields
\begin{equation}
X_{\phi_i} = -\{ \phi_i, \ \},
\qquad
i=1,\ldots, {\rm dim}( N_0),
\label{4.1}\end{equation}
of some functions $\phi_i$ on $M$, which are
the components of a momentum map $\phi$.
Suppose also that $N_0$ is a subgroup of a connected  Lie group $N$,
which acts on $M$ as a group of Poisson maps.
That is the action of $N$ on $M$ is a symmetry of $\{\ ,\ \}$, but it
is not necessarily  generated by a momentum map, except for the
sugroup $N_0$.
In practice this means that the action of $N$ is not required to
preserve the symplectic leaves in $M$.

Impose  now {\em first class constraints} of the form
\begin{equation}
\phi_i = c_i, \quad \hbox{$c_i$ some constant},
\label{4.2}\end{equation} on the
symmetry generators of the $N_0$-action,
that is fix the value of the momentum map to an infinitesimal
character of the Lie  algebra of $N_0$.
The constrained
manifold $M_c\subset M$ is given by
\begin{equation}
M_c:= \{\, p\in M\,\vert\, \phi_i(p)=c_i\,\}.
\label{4.3}\end{equation}
The constraints being first class means that $M_c$ is mapped to
itself by the action of $N_0$ on $M$.

Let us now make the additional assumption that $M_c\subset M$ is mapped
to
itself by the action of $N$ on $M$.
Then we can talk about the space of $N$-orbits in $M_c$ denoted as
\begin{equation}
M_{\rm red}:=M_c/N.
\label{4.4}\end{equation}
In a sufficiently regular situation, $M_{\rm red}$ is a manifold and
the space of smooth functions on $M_{\rm red}$ can be identified
with the space of smooth $N$-invariant functions on $M_c$.
If the embedding of $M_c$ in $M$ and
the action of $N$ on $M_c$ meet certain regularity conditions,
the following proposition is verified in a  standard manner.

\medskip\noindent
{\bf Proposition I.}
{\em
The Poisson bracket $\{\ , \ \}$ on $M$ induces a  Poisson bracket
$\{\ ,\ \}_{\rm red}$ on $M_{\rm red}$.
The induced Poisson bracket is determined  by the
extension-computation-restriction algorithm.
That is if $f$ and $g$ are $N$-invariant functions on $M_c$ then
the $N$-invariant function $\{ f, g\}_{\rm red}$ on $M_c$ is given by
\begin{equation}
\{ f, g\}_{\rm red}(p):=\{ \tilde f, \tilde g\}(p) \qquad \forall p\in
M_c,
\label{4.5}\end{equation}
where $\tilde f$, $\tilde g$ are  arbitrary (locally defined, smooth)
extensions of $f$ and $g$ to $M$.
}\medskip

If $N_0$ is the trivial group, the above
reduction procedure is just standard {\em Poisson reduction}
defined by factoring $M=M_c$ by the group of Poisson maps $N$.
The other extreme case is $N=N_0$,  when it becomes
{\em reduction by first class constraints},
which can  be also viewed as Marsdein-Weinstein reduction
applied to the symplectic leaves in $M$.
The general case interpolates between these two well-known
reduction procedures, and for this reason
we will refer to it by means of the term ``{\it hybrid reduction}''.

\medskip
We can specialize the general notion of hybrid reduction
to recover the KdV situation as follows.
First, the role of $(M, \{\ ,\ \})$ is played by $(\M,\{\ ,\ \}_\tau)$,
where the manifold $\M$ is defined in (\ref{3.20}) and the Poisson
bracket $\{\ ,\ \}_\tau$ on $\M$,
 the restriction of (\ref{3.18}), can be written
\begin{equation}
\{ f, g\}_\tau =\int dx\,
\left\langle J_0,
\left[ {\delta f\over \delta J_0} ,
 {\delta g\over \delta J_0} \right]\right\rangle
-\left\langle  {\delta f\over \delta J_0},
\pa_x {\delta g\over \delta J_0}\right\rangle
-\left\langle J_{>0}^{<k}+\Lambda_{>0},
\left[ {\delta f\over \delta J_{>0}^{<k}},
 {\delta g\over \delta J_{>0}^{<k}} \right]\right\rangle,
\label{4.6}\end{equation}
where ${\delta f\over \delta J_0(x)}\in \A_0$ and
${\delta f\over \delta J^{<k}_{>0}(x)}\in \A^{>-k}_{<0}$ for any
smooth function $f$ on $\M$.
The role of the group $N_0$ is played by the group $\N_0$
whose Lie algebra is by definition
$\tilde \A_0^{\leq -k}:=\{ \alpha\,\vert\, \alpha(x)\in \A_0^{\leq
-k}\}$,
with appropriate  smoothness and boundary conditions for $\alpha(x)$.
The group $\N_0$ acts on the manifold $\M$ via
\begin{equation}
e^\alpha: (\pa_x+J_0 +J_{>0}^{<k} +\Lambda_{>0})
\mapsto
e^\alpha (\pa_x+J_0 +J_{>0}^{<k} +\Lambda_{>0})  e^{-\alpha}
= e^\alpha (\pa_x+J_0) e^{-\alpha} +J_{>0}^{<k} +\Lambda_{>0},
\label{4.7}\end{equation}
where the last eqality follows from the relation
$\A_{>0}^{<0}=\{0\}$, which is a consequence of ({3.6}).
Combining (4.7) and (4.6) we see that the action of $\N_0$ on $\M$
is indeed a Hamiltonian action. The corresponding momentum
map
$\phi: \M \rightarrow \left(\tilde \A_0^{\leq -k}\right)^*\simeq
\tilde \A_0^{\geq k}:=\{ \beta\,\vert\, \beta(x)\in \A_0^{\geq k}\}$
is given by
\begin{equation}
\phi: \L=(\pa_x + J_0 +J_{>0}^{<k} +\Lambda_{>0})\mapsto J_0^{\geq k}
\quad\hbox{with}\quad J_0=J_0^{\geq k}+J_0^{<k}.
\label{4.8}\end{equation}
We define first class constraints by setting the
current component $J_0^{\geq k}$ generating the $\N_0$ action
equal to the constant $\Lambda_0\in \A_0^k$. This  leads to the
constrained manifold  $\M_c\subset \M$,
\begin{equation}
\M_c=\left\{ \L=\pa_x + J^{<k}_0(x) +J_{>0}^{<k}(x)+\Lambda\,\vert\,
J^{<k}_0(x)\in\A^{<k}_0,\,\,\, J_{>0}^{<k}(x)\in  \A_{>0}^{<k}
\,\right\}.
\label{4.9}\end{equation}
By the identification $j:=J_0^{<k} + J_{>0}^{<k}$,
$\M_c$ in (\ref{4.9})  the same as the constrained manifold
defined  in (\ref{3.8}).
Finally, it is clear that the role of the group $N$ in the
hybrid reduction is played by the group $\N$ acting  on
$\M$ according to
\begin{equation}
e^\alpha: \L\mapsto e^{{\rm ad} \alpha}\left(\L\right)=
e^\alpha \L e^{-\alpha},
\qquad
\forall\,\L\in \M,
\quad
\alpha(x)\in \A_0^{<0}.
\label{4.10}\end{equation}
This is a Poisson action\footnote{Equation (\ref{4.10}) defines
a Hamiltonian action of $\N$ on $\M$,
with momentum map $J_0\mapsto J_0^{> 0}$, if $\A_{>0}^{<k}=\{ 0\}$.}
preserving $\M_c\subset \M$.
Since the action of $\N$ on $\M_c$ admits DS gauges,
$\M_{\rm red}=\M_c/\N\simeq \M_V$ is a manifold and the
induced Poisson  bracket $\{\ ,\ \}_{\rm red}$
of Proposition I is just the ``second'' Poisson
bracket of the KdV system given in (\ref{3.16}).

The hybrid reduction interpretation is useful not only for
understanding the origin of the ``second'' Poisson bracket, but
it also sheds light on the commuting Hamiltonians of the KdV system.
Namely, the commuting Hamiltonians can be explained combining this
reduction procedure with the r-matrix (Adler-Kostant-Symes) scheme,
which tells us that  the monodromy invariants of $\L$ provide
commuting Hamiltonians on $\M$.

\medskip\noindent  {\bf Remarks.} {\em a)}
If $\left[\A_0^{<0}, \Lambda_{>0}+ \A_{>0}^{<k} \right]\neq \{ 0\}$
then the action of $\N$ on $\M$ is  not a  Hamiltonian action.
This  is the case in examples
for which  $\Lambda$ is not a
semisimple element of minimal positive $d_\sigma$-grade.

\noindent {\em  b)}
If  the $d_\sigma$-grade $k$ of $\Lambda$  is
large enough then $\Lambda_0=0$ and $\N_0$ is trivial.
In this case there are no first class constraints, $\M_c=\M$,
and we are dealing with Poisson reduction.

\noindent {\em c)}
If   $k=1$ then $\N_0=\N$
 and   the hybrid reduction
becomes reduction by first class constraints.

\noindent
{\em d)} The space
$(\tilde \A, \{\ ,\ \}_\tau)$ can be naturally embedded as a
Poisson submanifold in the dual of a Lie algebra
having a Lie-Poisson bracket engendered by an r-matrix (see
\cite{RSTS}).
In the periodic case, the Lie algebra in question is given by
$\ell(\widehat{\G},\hat \tau)\subset \hat \G\otimes
{\bf C}[\lambda,\lambda^{-1}]$, where $\widehat{\G}$ is the central
extension of the loop algebra $C^\infty(S^1,\G)$,
and $\hat\tau$ is the automorphism of $\hat \G$ naturally
induced from the automorphism $\tilde \tau$ of $\G$
that corresponds to the grading $d_\tau$ of $\A$
by Kac's theorem \cite{kac}.
$\A$ can be realized as  $\A=\ell(\G,\tilde \tau)$
and $d_\tau$ becomes the homegeneous grading in this realization.
The action of $\hat \tau$ on $\hat \G$ is defined by  extending
the action of $\tilde\tau$ to  $C^\infty(S^1,\G)$ in a pointwise fashion
and acting as the identity on the centre of $\hat \G$.
The r-matrix corresponds to the splitting of
$\ell(\widehat{\G},\hat \tau)$ according to negative and non-negative
powers of $\lambda$.
Observe that the two loop parameters $x\in S^1$ and
$\lambda$ play completely different roles in the construction.

\medskip
\section{$\cal W$-algebras related to KdV systems}
\setcounter{equation}{0}

Notice  from formula (\ref{4.6}) of the Poisson bracket
$\{\ ,\ \}_\tau$ on $\M$ that the component $J_0$ of $J=(J_0+J_{>0})$
satisfies the standard current algebra Poisson bracket associated
to the reductive Lie algebra $\A_0$,  with
the central extension being defined by means of the restriction of
the invariant scalar product of $\A$ to $\A_0$.
Going to the constrained manifold $\M_c$, this ensures that we can
talk about the Poisson subalgebra of the gauge invariant functions
depending only on the component $j_0$ of the constrained current
$j=(j_0+j_{>0})$.
In particular, we can talk about the Poisson bracket algebra of the
generators of the subring ${\cal R}_0\subset {\cal R}$
of  gauge invariant differential polynomials depending only on $j_0$.
Following \cite{SdeC}, next we present an additional condition
on the KdV  quadruplet $(\A, d_\sigma, \Lambda, d_\tau)$,  which
ensures that  ${\cal R}_0$ is a finitely generated,
freely generated differential ring and the corresponding subalgebra
of the ``second'' Poisson bracket algebra is a $\W$-algebra of the
standard type.

Using the decomposition $\Lambda=\Lambda_0 +\Lambda_{>0}$,
suppose that $\Lambda_0\neq 0$
and the  KdV quadruplet $(\A, d_\sigma, \Lambda, d_\tau)$
satisfies the ``strong non-degeneracy condition''
\begin{equation}
\kerr\cap \A_0^{<0}=\{0\}.
\label{5.1}\end{equation}
This implies the original non-degeneracy condition  in (\ref{3.6}).
If condition (5.1)  is satified,
we can choose the complementary space $V$ in (\ref{3.10})
so that
\begin{equation}
V=V_0 + V_{>0}
\quad\hbox{with}\quad
\A_0^{<k}=[\Lambda_0, \A_0^{<0}] + V_0,
\quad
V_{>0}=\A_{>0}^{<k}.
\label{5.2}\end{equation}
Note that $V_0\subset \A_0^{\leq 0}$ since
$[\Lambda_0, \A_0^{<0}\cap \A_0^{>-k}]=\A_0^{>0}\cap \A_0^{<k}$
on account of (\ref{5.1}).
The action of the gauge group $\N$  on $\M_c$ given in (\ref{3.9})
can be written
 \begin{equation}
e^\alpha: j_0 \mapsto
e^\alpha(\pa_x+ j_0 + \Lambda_0)e^{-\alpha} -\Lambda_0-\pa_x,
\quad
j_{>0} \mapsto e^\alpha(j_{>0} + \Lambda_{>0})e^{-\alpha},
\quad
\forall\, \alpha(x)\in \A_0^{<0},
\label{5.3}\end{equation}
so it does not explicitly mix  the components  $j_0$ and $j_{>0}$ of
$j=(j_0+j_{>0})$.
Combining  (\ref{5.1}), (\ref{5.2}) and (\ref{5.3}),
we see that the DS gauge fixing
can be performed purely at the $d_\tau$-grade zero level.
This means that the gauge transformation $e^\alpha$ that brings
$j=(j_0+j_{>0})$ to the DS
normal form $j_V=j_{V_0} + j_{V_{>0}}$ corresponding
to $V$ in (\ref{5.2}) depends only on $j_0$, $\alpha=\alpha(j_0)$.
As a consequence, the ring $\cal R$ of gauge invariant
differential polynomials in $j$ is generated by the components of
$j_{V_0}=j_{V_0}(j_0)$ and $j_{V_{>0}}=j_{V_{>0}}(j_{>0}, j_0)$.
This immediately implies the first statement of the following
proposition.

\medskip\noindent
{\bf Proposition II.}
{\em Suppose that $\Lambda_0\neq 0$
and  (\ref{5.1}) is satisfied.
Then the  subring ${\cal R}_0\subset {\cal R}$ of
$\N$-invariant differential
polynomials in $j_0$ is freely generated by the components of
$j_{V_0}(j_0)$ for any DS normal form $j_V=j_{V_0}+j_{V_{>0}}$
associated to a decomposition of the type in (\ref{5.2}).
There exists a generating set of ${\cal R}_0$
whose induced Poisson brackets satisfy the
$\W^{\A_0}_\S$-algebra for  the $sl_2$ embedding
$\S={\rm span}\{ I_-, I_0, I_+\}\subset \A_0$,
$[I_0, I_\pm] = I_\pm$, $[I_+, I_-]=2I_0$, defined  by $I_+:=\Lambda_0$.
}
\medskip

\noindent{\em Proof.}
The relation  $d_\sigma(\Lambda_0)=k\Lambda_0$ implies that
$\Lambda_0$ is a nilpotent element of $\A_0$
and thus $I_+:=\Lambda_0$  determines  an $sl_2$
subalgebra  $\S\subset \A_0$ which is unique up to conjugation
\cite{jac}.
The generator $I_-$ of $\S\subset \A_0$ can be chosen so as to also
have  $d_\sigma(I_-)=-kI_-$.
With this $I_-$,  define
\begin{equation}
V_0:=\kerrr.
\label{5.4}\end{equation}
It is not hard to verify that  $V_0$ in (\ref{5.4})
 satisfies  (\ref{5.2}).
By the definition of $\W_\S^{\A_0}$ \cite{bais,rep},
the  components of  $j_{V_0}$
 generate the  $\W_\S^{\A_0}$-algebra {\em with respect to
the Dirac bracket} determined by the second class
constraints that reduce  the $\A_0$ valued
current $J_0$ to $(j_{V_0}+\Lambda_0)$.
We wish  to show that the Dirac brackets of the components
of $j_{V_0}$ coincide with the induced Poisson brackets
of the ${\cal R}_0$ generators
given by  the components of $j_{V_0}(j_0)$.
For this, let us  consider the subalgebra $\Gamma\subset \A_0^{<0}$,
\begin{equation}
\Gamma:=\A_0^{<-{k/2}} + {\cal P}^{-k/2},
\label{5.5}\end{equation}
where  ${\cal P}^{-k/2}$ is defined
to be\footnote{We are using here a version of
``symplectic halving''  \cite{rep}.
If $k$ is odd then ${\cal P}^{-k/2}=\{ 0\}$.}
a maximal subspace of
$\A_0^{-k/2}$ for which $\langle \Lambda_0, [X, Y]\rangle =0$ for any
$X, Y\in {\cal P}^{-k/2}$.
It can be checked  that the
 constraints given by
\begin{equation}
\phi_i(J_0)=\langle \gamma_i, J_0 -\Lambda_0\rangle =0,
\label{5.6}\end{equation}
where $\{ \gamma_i\}$ is a basis of $\Gamma$,
are a set of {\em first class constraints} on $J_0$,
 which upon gauge fixing  lead
to {\em the same}  second class constraints that restrict
$J_0$ to have the form of $(j_{V_0} + \Lambda_0)$.
Indeed, the $\Gamma$-constrained current, $J_0^\Gamma$,  has
the form
\begin{equation}
J_0^\Gamma=(j_0^\Gamma +\Lambda_0)
\quad\hbox{with}\quad j_0^\Gamma(x)\in \Gamma^\perp,
\label{5.7}\end{equation}
where $\Gamma^\perp\subset \A_0$ is the annihilator of $\Gamma$
with respect to the scalar product $\langle\ ,\ \rangle$,
and we have
\begin{equation}
\Gamma^\perp=[\Lambda_0, \Gamma] + V_0
\label{5.8}\end{equation}
with {\em the  same} complementary space $V_0$ as in (\ref{5.2}).
Then we can use the fact  that $j_{V_0}(j_0)$
commutes with the first class constraints
associated to $\Gamma$, as $\Gamma\subset \A_0^{<0}$,
 to conclude  from the standard formula for the Dirac bracket
that the Dirac brackets of the components of
$j_{V_0}(j_0)$  coincide with their induced Poisson brackets
determined by hybrid reduction.
{\em Q.E.D.}
\medskip

Proposition II is due to Miramontes et al \cite{SdeC}.
We note that the symplectic halving part of the above
proof is not contained  in \cite{SdeC}, where
the case of $\A=\G\otimes {\bf C}[\lambda, \lambda^{-1}]$
was considered.

\medskip
\noindent
{\bf Remarks.}
{\em e)} Under the strong non-degeneracy condition,
the components
of $j_{V_{>0}}(j_{>0}, j_0)$ {\em commute} with those of $j_{V_0}(j_0)$
with respect to  the ``second'' Poisson bracket \cite{MiPo}.

\noindent
{\em f)}  If $\Lambda$ has $d_\sigma$-grade one
 then $[\Lambda_{>0}, \A_0^{<0}]\subset \A_{>0}^{<1}=\{0\}$,
and so  the non-degeneracy condition in (\ref{3.6})
is equivalent to  (\ref{5.1}).
In this case  the ``second'' Poisson
bracket algebra of the KdV system
associated to the quadruplet $(\A,d_\sigma, \Lambda, d_\tau)$
is just the $\W_\S^{\A_0}$-algebra.

\noindent
{\em g)} The strong non-degeneracy condition is a sufficient but
{\em not} a necessary condition for the presence of a $\W$-subalgebra
in the second Poisson bracket algebra of a KdV system  \cite{MiPo}.
An instructive  set of examples
 is obtained  by taking
$\A=sl_n \otimes {\bf C}[\lambda, \lambda^{-1}]$,
$d_\sigma$: the principal grading, $d_\tau$: the homogeneous
grading, and $\Lambda:= (\Lambda_n)^l$ for some
$1<l<n$  {\em relatively prime} to $n$.
Here $\Lambda_n:=\left(\lambda e_{n,1} + \sum_{i=1}^{n-1} e_{i,
i+1}\right)$
is the grade one regular element from the principal Heisenberg
subalgebra.
For these examples (\ref{3.6})
holds but the strong non-degeneracy condition (5.1)
is not satisfied unless $l=2$ and $n$ is odd.
Nevertheless, the presence of a
$\W_\S^{\A_0}$-subalgebra
can  be exhibited in the second Poisson bracket algebra \cite{DF+}.
The case $l=(n-1)$ is treated in \cite{MiPo} by a different method.

\noindent
{\em h)}  In the general case, it is known \cite{MiPo}
that if $\Lambda_0\neq 0$ then
the ``second'' Poisson bracket of the KdV system
$(\A, d_\sigma, \Lambda, d_\tau)$
enjoys a conformal
invariance property given by a (non-unique) Poisson action of
the conformal group on the phase space.
The existence of a corresponding Virasoro density,
not to mention a --- standard or new --- $\W$-algebra,
is not clear at present.

\medskip
\section{Classification of mod-KdV and KdV type systems}
\setcounter{equation}{0}

We have seen that a modified KdV type system can be associated
to any triplet $(\A, d_\sigma, \Lambda)$  and that a KdV type system
can be associated to
any  quadruplet $(\A, d_\sigma, \Lambda, d_\tau)$
subject to  $\tau\preceq \sigma$,
$\A_0^{<0}\neq \{0\}$ and the non-degeneracy condition in (\ref{3.6}).
The classification of modified KdV type (KdV type) systems requires
listing
all triplets (quadruplets) modulo an  equivalence relation taking care
of
possible isomorphisms  of hierarchies corresponding to different
triplets
(quadruplets).
We are  very  far from having a complete classification,
but  some progress  has been made.

In \cite{McI}  various conditions on the gradings
admitting  a {\em grade one} semisimple element were described.
It was explained that $\sigma_i\in \{ 0,1\}$ for all $i=0,1,\ldots, r$
is a necessary condition for the grading $d_\sigma$ in (\ref{2.6})
to admit a grade one semisimple element.
A  sufficient  condition (which was conjectured to be necessary  too)
was also given.
The sufficient condition of \cite{McI} is a constructive one, and
for instance  in the case of  $\A=sl_n\otimes {\bf C}[\lambda,
\lambda^{-1}]$
it is satisfied for any choice of $\sigma_i\in \{0,1\}$.
The question of  possible isomorphisms of modified KdV systems
associated to different triplets was also studied in \cite{McI},
and the interested reader is urged  to consult this reference.

The series of papers in [13--17]
contains the assumption  that $\Lambda\in \A$ belongs to
a graded Heisenberg subalgebra
(a maximal abelian subalgebra of semisimple elements) of $\A$
and, except in \cite{Prin3},  it was also assumed that
$\A$ is a non-twisted loop algebra.
Although these assumptions are not essential, as we have seen,
it would be important  to know whether every graded semisimple
element is contained in a graded Heisenberg subalgebra.
If the answer is positive, then the  known classification  of the
inequivalent graded Heisenberg subalgebras
  of the affine Lie algebras,
which is described
in \cite{KP} in the non-twisted case,
would yield  a partial classification of the
generalized  (modified) KdV systems.
This classification would still be incomplete for  several
reasons, including the fact that
in general there exist different  gradings of $\A$ that reduce
to the  grading of a given Heisenberg subalgebra of $\A$,
which itself is unique up to a constant.

Since the description of  all  possible triplets $(\A,d_\sigma,
\Lambda)$
is a prerequisite for finding the list of the quadruplets
$(\A, d_\sigma, \Lambda, d_\tau)$,
we know  even less about the latter problem.
As for the possible auxiliary gradings $d_\tau$ admitted for a
given triplet $(\A, d_\sigma, \Lambda)$, no general result is available.
There is no difficulty in choosing $d_\tau$ so that
$\tau\preceq \sigma$ and $\A_0^{<0}\neq \{0\}$,
but the non-degeneracy condition in (\ref{3.6}) places a
non-trivial restriction on $d_\tau$.
However, this restriction disappears in the important special
case for which the graded semisimple element $\Lambda$ is {\em regular}.
By definition, for $\Lambda$ a {\em regular} semisimple element
$\ker$ is  an {\em abelian} subalgebra of $\A$
(a graded Heisenberg subalgebra if  $\Lambda$ has definite grade).
In the regular case  $\ker$   consists of semisimple
elements, and thus it cannot intersect
 $\A_0^{<0}$, which  consists of nilpotent elements.

In the regular case more classification results
are available than in the general case.
We now  explain how this comes about concentrating, for simplicity,
on the non-twisted affine algebras.
Let $\tilde \sigma$ denote the finite order
inner automorphism of  $\G$ corresponding \cite{kac}
to the grading $d_\sigma$ of
$\A=\G\otimes {\bf C}[\lambda, \lambda^{-1}]$.
If  $\Lambda(\lambda)\in \A$
is a semisimple element of definite $d_\sigma$-grade, then
its ``projection''  $E\in \G$
--- $E\in \G$ is obtained from $\Lambda(\lambda)\in \A$
by replacing the formal parameter
$\lambda$ by $1$,  that is  $E:=\Lambda(\lambda=1)$ ---
is a semisimple eigenvector of $\tilde \sigma$.
Moreover, $\Lambda(\lambda)\in \A$ is regular if and only
if $E\in \G$ is regular in the usual finite dimensional sense.
But if $E$ is a regular semisimple eigenvector of $\tilde \sigma$,
then $\tilde \sigma$ preserves the
Cartan subalgebra
$\H\subset \G$ defined by $\H:=\kerh$,
and thus it gives rise to an
element of the Weyl group ${\bf W}(\G)$ acting on $\H$.
On the other hand, the conjugacy classes in ${\bf W}(\G)$
whose representatives admit a regular eigenvector
--- themselves called regular conjugacy classes ---
are all known \cite{Sp}.
Reversing  the above arguments, all  of the
graded regular semisimple elements of $\A$ can be
constructed\footnote{Using arbitrary Cartan preserving
automorphisms of $\G$ of
finite order instead of just inner ones,
an analogous result applies to general  affine Lie algebras.}
out of the regular eigenvectors
of representatives of the regular conjugacy classes in  ${\bf W}(\G)$.
There is a  ``lifting ambiguity'' involved in the construction
since  non-conjugate inner automorphisms of $\G$
may give rise  to the same Weyl transformation.
This ambiguity has not been settled yet,
but a particularly nice lift is given by the following result.

\medskip\noindent
{\bf Proposition III.} {\em
Let $w\in {\bf W}(\G)$ be a representative
of a regular  conjugacy class of the Weyl group
acting on the Cartan subalgebra $\H\subset \G$.
Then there exists a finite order automorphism
$\hat w=\exp\left(2i\pi {\rm ad} I_0/m\right)$ of $\G$
that reduces  to $w$ on $\H$, where $m$ is the
order of $w$ and the following statements are valid:

\smallskip
\noindent i)  $I_0\in \G$ is the ``defining vector'' \cite{dyn} of an
$sl_2$ subalgebra of $\G$.

\smallskip
\noindent ii) The largest  eigenvalue   $j_{\rm max}$
of ${\rm ad} I_0$ equals  $(m-1)$ except for some
cases in $F_4$ and $E_{6,7,8}$.

\smallskip
\noindent iii) The order of $\hat w$ is $m$ (resp.~$2m$)
if ${\rm ad} I_0$ has only integral (resp.~also half-integral)
 eigenvalues.

\smallskip
\noindent iv)
$\hat w$ has  a regular eigenvector
$E=(C_1 + C_{-(m-1)})\in \H$  with  eigenvalue
$e^{2i\pi/m}$,  $[I_0, C_l]=lC_l$. 

\smallskip
\noindent v)
For $I_+:=C_1$ appearing in  $E$ above,
there exists $I_-\in \G$ so that $[I_0, I_\pm ]=I_\pm$, $[I_+,
I_-]=2I_0$.}

\medskip
Proposition III generalizes the celebrated
relation  between the Coxeter class in  ${\bf W}(\G)$ and
the principal $sl_2$ subalgebra of $\G$ due to Kostant \cite{Kost}.
It was verified in \cite{DF} putting together
isolated results of \cite{Sp,leur,bouw}.
The  $sl_2$ embeddings associated by Proposition III to the regular
conjugacy classes in the Weyl group are  given explicitly
in \cite{DF} in terms of the standard
classifications of the $sl_2$ subalgebras of  $\G$  \cite{dyn}
and the conjugacy classes of ${\bf W}(\G)$ \cite{cart}.

Finally, we wish to mention an application of Proposition III to KdV
systems.
Define $\Lambda \in \A=\G\otimes {\bf C}[\lambda,\lambda^{-1}]$
by $\Lambda:=(C_1 + \lambda C_{-(m-1)})$ using
$E=\Lambda(\lambda=1)$ occurring in  property iv).
It follows that $\Lambda$ is a  regular semisimple element
of grade $\nu$ with respect to
$d_\sigma:=\nu \left(m\lambda {d\over d\lambda} + {\rm ad I_0}\right)$,
where
$\nu\in \{ 1,2\}$ equals the order of $\hat w$ divided by the order of
$w$.
In fact,  the homogeneous grading $d_\tau:=\lambda {d\over d\lambda}$
satisfies the relation $\tau \preceq \sigma$ if and only if
$j_{\rm max}=(m-1)$.
(In the exceptional cases for which this does not
hold one has $j_{\rm max}=m$.)
This means that in the cases for which  $j_{\rm max}=(m-1)$,
the quadruplet $(\A, d_\sigma, \Lambda, d_\tau)$
can be used to construct a KdV type system.
Then $\A_0=\G$ and the strong non-degeneracy condition
of (\ref{5.1}) is satisfied as a consequence of
$d_\sigma$ being defined in terms of the $sl_2$ generator
 $I_0$  and the equality $\Lambda_0=I_+$ by property v).
The subring ${\cal R}_0$ mentioned in Proposition II exhausts $\cal R$
for these cases, and the second Hamiltonian structure
of the corresponding KdV system is just the
$\W_\S^\G$-algebra for the $sl_2$ embedding $\S$ defined by $I_0$.
A detailed investigation of these systems, focused
on developing  a Gelfand-Dickey type description analogous
to that of the $n$-KdV system in (\ref{1.1}),
is given  in \cite{DF,FM}.
For $\G$ a classical Lie algebra,
a  Gelfand-Dickey type pseudo-differential operator model
was found for about half of these systems,
which were shown to also arise from reductions  of KdV systems
associated
to $gl_n$ by means of involutive discrete symmetries.
The interested reader may consult the papers in \cite{DF,FM} for
details.

\bigskip

\bigskip

\noindent
{\large \bf  Acknowledgements.}
I wish to thank Francois Delduc,
Carlos Fern\'andez-Pousa,
Ian Marshall, Luis Miramontes, Izumi Tsutsui
and George Wilson for useful discussions and comments.


\end{document}